\documentclass[conference]{IEEEtran}
\IEEEoverridecommandlockouts

\newcommand{\reffig}[1]{Fig.~\ref{#1}}
\newcommand{\reftab}[1]{Table~\ref{#1}}
\newcommand{\refsec}[1]{Section~\ref{#1}}
\newcommand{\refalgo}[1]{Algorithm~\ref{#1}}

\usepackage{adjustbox}

\usepackage[linesnumbered,ruled,vlined]{algorithm2e}
\SetKwInput{KwInput}{Input}
\SetKwBlock{Repeat}{repeat}{}
\SetKwInOut{Initialization}{Initialization}
\usepackage{amsmath}
\usepackage{bm}
\usepackage{url}
\usepackage{balance}
\usepackage{etoolbox}
\makeatletter
\patchcmd{\@makecaption}
  {\scshape}
  {}
  {}
  {}
\makeatother

\usepackage[numbers,square, comma, sort&compress]{natbib}
\usepackage{tabularx,booktabs}
\usepackage{pict2e}
\usepackage[utf8]{inputenc}
\usepackage[T1]{fontenc}
\usepackage{slashbox}
\usepackage{multirow}
\usepackage{adjustbox}
\usepackage{lipsum,multicol}
\usepackage{pifont}
\newcommand{\cmark}{\ding{51}}%

\usepackage{xcolor}
\usepackage{enumitem}
\usepackage{framed}
\usepackage{listings}
\definecolor{azuremist}{rgb}{0.94, 1.0, 1.0}
\definecolor{black}{RGB}{0,0,0}
\definecolor{gray}{RGB}{102,102,102}        
\definecolor{function}{RGB}{0,102,153}      
\definecolor{lightgreen}{RGB}{102,153,0}    
\definecolor{bluegreen}{RGB}{51,153,126}    
\definecolor{lightbluegreen}{RGB}{11,103,70}    
\definecolor{magenta}{RGB}{217,74,122}  
\definecolor{orange}{RGB}{226,102,26}       
\definecolor{purple}{RGB}{125,71,147}       
\definecolor{green}{RGB}{113,138,98}        
\lstdefinelanguage{SQL}{
  showstringspaces=false,
  xleftmargin=5.0ex,
  firstnumber=1,
  numberstyle=\small\tt\color{gray},
  tabsize=1,
  numbers=left,
  morekeywords = [3]{SELECT,select,WHERE,where,OPTIONAL},
  morekeywords = [4]{},
  morekeywords = [5]{},
  morekeywords = [6]{+,=,:=,.,;,,,-,!,=,~,>,<},
  morekeywords = [7]{},
  keywordstyle = [3]\color{bluegreen},
  keywordstyle = [4]\color{lightgreen},
  keywordstyle = [5]\color{magenta},
  keywordstyle = [6]\color{orange},
  keywordstyle = [7]\color{function},
  sensitive = true,
  morecomment = [l][\color{gray}]{--},
  morecomment = [s][\color{gray}]{/*}{*/},
  morestring = [b][\color{purple}]",
  morestring = [b][\color{purple}]'
}

\usepackage[normalem]{ulem}

\usepackage{subfigure}
\usepackage{url}
\usepackage{amsmath,amssymb,amsfonts}
\usepackage{algorithmic}
\usepackage{graphicx}
\usepackage{textcomp}

\def\BibTeX{{\rm B\kern-.05em{\sc i\kern-.025em b}\kern-.08em
    T\kern-.1667em\lower.7ex\hbox{E}\kern-.125emX}}
\begin{document}

\title{SymphonyDB: A Polyglot Model for \\Knowledge Graph Query Processing}

\author{\IEEEauthorblockN{Masoud Salehpour}
\IEEEauthorblockA{University of Sydney}
\and
\IEEEauthorblockN{Joseph G. Davis}
\IEEEauthorblockA{University of Sydney}}

\maketitle

\begin{abstract}
Unlocking the full potential of Knowledge Graphs (KGs) to enable or enhance various semantic and other applications requires Data Management Systems (DMSs) to efficiently store and process the content of KGs. However, the increases in the size and variety of KG datasets as well as the growing diversity of KG queries pose efficiency challenges for the current generation of DMSs to the extent that the performance of representative DMSs tends to vary significantly across diverse query types and no single platform dominates performance. We present our extensible prototype, SymphonyDB, as an approach to addressing this problem based on a polyglot model of query processing as part of a multi-database system supported by a unified access layer that can analyze/translate individual queries just-in-time and match each to the likely best-performing DMS among Virtuoso, Blazegraph, RDF-3X, and MongoDB as representative DMSs that are included in our prototype at this time. The results of our experiments with the prototype over well-known KG benchmark datasets and queries point to the efficiency and consistency of its performance across different query types and datasets.

\end{abstract}


\section{Introduction}
\label{sec::introduction}
Knowledge Graphs (KGs) are large-scale collections of facts to represent real-world \textit{entities} and their \textit{interconnections}. KGs have gained widespread use in different domains including computer science, medicine, bioinformatics, education, and biology, among others~\cite{Refinement}. Several KGs such as Wikidata, YAGO, and Bio2RDF, to name a few, are openly available. As well, many private organizations such as Amazon have created KGs for different purposes such as \textit{similarity analysis}. The widespread adoption of KGs has highlighted the need for employing efficient Data Management Systems (DMSs). However, the rich diversity in \textit{variety} and \textit{size} of KG content pose challenges to DMSs to \textit{store} and \textit{query} KGs~\cite{Growing3,Growing2}. The individual \textit{queries} and applications executed on these systems have also become \textit{highly diverse}. These have been addressed in part through the development of a range of DMSs such as \textit{columnar-} and \textit{graph-based} stores. The consensus appears to be that a single, \textit{one-size-fits-all} DMS is unlikely to emerge for efficient KG query processing~\cite{watdiv}.

We develop a solution to this problem that is inspired by Ashby's First Law of Cybernetics~\cite{ashby} and Stonebraker et. al.~\cite{onesize2} which can be paraphrased in this context to state that the variety in the solution architecture should be greater than or at least equal to that of the variety displayed by the data and the queries. The \textit{requisite variety} is to be achieved through an architecture based on \textit{polyglot} model of query processing and access languages supported by a design that can analyze individual queries and match each to the likely best-performing database engine. From a conceptual standpoint, genuine polygloty will imply the requirement of employing multiple DMSs and friction-free translation across the different query and access languages at the polyglot access layer in order to optimize query execution performance. However, a review of the extant research and practitioner literature such as~\cite{fowler} reveals that polygloty has typically been interpreted as using different physical data stores depending on the type of application. For instance, traditional relational DMSs for financial data, document-stores for product catalog data, key-value stores for user activity logs, etc. However, it is not difficult to see that these interpretations typically suffer from lack of integration across the entire data which may lead to \textit{balkanized} data islands. 

Supporting multiple data models against a single, integrated backend can potentially address the growing requirements for performance~\cite{lookliu}. Based on this, over the last few years, there has been growing interest in employing multiple DMSs for query processing. This interest led to the development of some open-source platforms such as Apache Drill\footnote{\url{https://drill.apache.org}} as well as some academic prototypes such as~\cite{bigdawg}. However, these solutions mainly focused on applications such as \textit{ETL}, \textit{machine learning}, \textit{stream processing}, \textit{OLAP}, data integration, etc., and somewhat less attention has been paid to efficient KG query processing which is the focus of this paper.

We present \textit{SymphonyDB}, a prototype that provides polyglot support for KG query processing. It is a multi-database approach supported by an access management layer to provide a unified query interface for accessing the underlying DMSs. This layer receives the incoming workloads in the form of SPARQL queries and routes each of them to one (or more than one) of the more likely to be efficiently matched DMSs. A suitable Just-In-Time (JIT) query translation may be needed if the underlying DMSs accept different query languages. Handling such a JIT translation process is the essential responsibility of the access layer. Currently, \textit{SymphonyDB} has included \textit{Virtuoso}, \textit{Blazegraph}, \textit{\mbox{RDF-3X}}, and \textit{\mbox{MongoDB}} as representative DMSs. SymphonyDB has the potential to achieve \textit{requisite variety}. 

\newpage

Our contributions include:

\begin{itemize}
 \item
 Presenting a prototype, \textit{SymphonyDB} that can match KG query requirements with the best combination of DMS and storage representation to achieve consistent query execution performance. 
 
 \item
 Experimental evaluation and comparative performance analysis of \textit{SymphonyDB} against representative single DMSs in supporting different archetypal KG query types. The source code, data, and other artifacts have been made available at \url{https://github.com/m-salehpour/SymphonyDB}.
 
 \end{itemize}


\begin{figure}[ht]
\centering\includegraphics[width=0.4\textwidth]{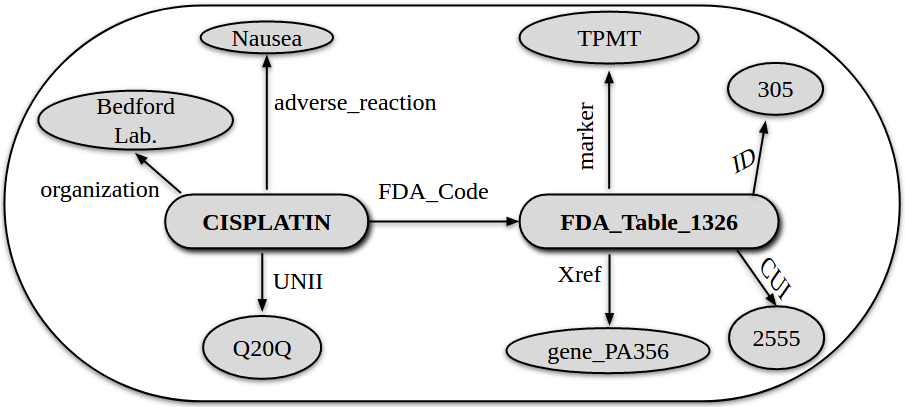}
\caption{An example of a simple Knowledge Graph (based on the LikedSPL KG).}
\label{fig::kg}
\end{figure}

\section{Background}
\label{sec::background}
 In this section, we present some preliminary information about the archetypal KG query types using a \textit{human-readable} example depicted in \reffig{fig::kg}. This shows a small extract from the LinkedSPL KG which includes all sections of FDA-approved prescriptions and over-the-counter drug package inserts from DailyMed. The content of this KG subset can be represented by the following RDF triples\footnote{The content of a KG usually represents a large \textit{set of triples} of the form $<$subject predicate object$>$ creating a graph~\cite{CharacteristicsRDF}}:

\begin{figure}[ht]
    \centering
    \begin{lstlisting}[language=SQL,basicstyle=\scriptsize\ttfamily,numbers=left,tabsize=2,showstringspaces=false,numberstyle=\scriptsize\tt\color{gray}]
    CISPLATIN     UNII               "Q20Q"
    CISPLATIN     adverse_reaction   "Nausea"
    CISPLATIN     organization       "Bedford Lab"
    CISPLATIN     FDA_Code           "Table_1326"
    Table_1326    marker             "TPMT"
    Table_1326    ID                  305
    Table_1326    CUI                 2555
    Table_1326    Xref               "gene_PA356"
    \end{lstlisting}
    \label{fig:triples-KG-fda}
\end{figure}

An example of a query\footnote{Queries are formulated in the form of SPARQL. In this paper, we assume that the reader is familiar with the basic concepts of querying KG, e.g., the SELECT clauses.} is given below. It asks for ``UNII'' of a given drug (i.e., ``CISPLATIN''). In this query, ``?unii'' is a variable to return the result which is ``Q20Q''.

\begin{figure}[ht]
    \centering
    \begin{lstlisting}[language=SQL,basicstyle=\ttfamily,numbers=left,tabsize=2,showstringspaces=false,numberstyle=\scriptsize\tt\color{gray}]
    SELECT ?unii
    WHERE {
    CISPLATIN UNII ?unii . }
    \end{lstlisting}
\end{figure}

 KG queries may contain a set of triple patterns such as \textit{\mbox{``CISPLATIN UNII ?unii''}} in which the subject, predicate, and/or object can be a variable. Each triple pattern typically returns a subgraph. This resultant subgraph can be further \textit{joined} with the results of other triple patterns to return the final resultset. In practice, there are three major types of join queries: (i) subject-subject joins (aka, star-like), (ii) subject-object joins (aka, chain-like or path), and (iii) tree-like (i.e., a combination of subject-subject and subject-object joins). These query types are explained below.

\textbf{Subject-subject joins.} A subject-subject join is performed by a DMS when a KG query has at least two triple patterns such that the predicate and object of each triple pattern is a given value (or a variable), but the subjects of both triple patterns are replaced by the \textit{same} variable. For example, the following query looks for all drugs for which their UNII and adverse reactions are equal to the given values (the result will be ``CISPLATIN'').

\begin{figure}[ht]
    \centering
    \begin{lstlisting}[language=SQL,basicstyle=\ttfamily,numbers=left,tabsize=2,showstringspaces=false,numberstyle=\scriptsize\tt\color{gray}]
    SELECT ?x
    WHERE {
    ?x UNII             "Q20Q"   .
    ?x adverse_reaction "Nausea" . }
    \end{lstlisting}
\end{figure}

\textbf{Subject-object joins.} A subject-object join is performed by a DMS when a KG query has at least two triple patterns such that the subject of one of the triple patterns and the object of the other triple pattern are replaced by the same variable. For example, the following query looks for all drugs that their CUI\footnote{CUI (aka, RxCUI) is a unique drug identifier} is equal to 2555 (``CISPLATIN'' is the result).

\begin{figure}[ht]
    \centering
    \begin{lstlisting}[language=SQL,basicstyle=\ttfamily,numbers=left,tabsize=2,showstringspaces=false,numberstyle=\scriptsize\tt\color{gray}]
    SELECT ?y
    WHERE {
    ?x  CUI       2555  .
    ?y  FDA_Code  ?x    . }
    \end{lstlisting}
\end{figure}

\textbf{Tree-like joins.} A tree-like join consists of a \textit{combination} of subject-subject and subject-object joins. For example, the following query looks for the ``Xref'' of all drugs with ``UNII'' of ``Q20Q'' that their ``CUI'' is equal to 2555 (the result will be ``gene\_PA356'').

\begin{figure}[ht]
    \centering
    \begin{lstlisting}[language=SQL,basicstyle=\ttfamily,numbers=left,tabsize=2,showstringspaces=false,numberstyle=\scriptsize\tt\color{gray}]
    SELECT ?y
    WHERE {
    ?x  UNII        "Q20Q"  .
    ?x  FDA_Code    ?z      .
    ?z  CUI         2555    .
    ?z  Xref        ?y      . }
    \end{lstlisting}
\end{figure}

In addition to the query types, we provide a brief explanation of query \textit{selectivity} and \textit{optional} patterns. The selectivity of a query is typically represented by the fraction of triples matching the query pattern. Based on this, each query type can typically be either \textit{high-selective} or \textit{low-selective}. A query can be considered as low-selective when a large number of triples (as compared to the total number of stored triples) needs to be scanned before returning the resultset~\cite{selectivity}. Moreover, as explained previously, queries return resultsets only when the entire query pattern matches the content of the KG. However, some queries may contain optional patterns to allow KG queries to return a resultset even if the optional part of the query is not matched since completeness and adherence of KG content to their formal ontology specification is not always enforced.

\section{SymphonyDB}
\label{sec::SymphonyDB}

   In this section, we present \textit{SymphonyDB}, a prototype that provides polyglot support for KG query processing. SymphonyDB is a multi-database solution supported by an access management layer. Currently, \textit{SymphonyDB} includes \textit{Virtuoso}, \textit{Blazegraph}, \textit{\mbox{RDF-3X}}, and \textit{\mbox{MongoDB}} as representative DMS types.
   An abstract overview of the interactions across DMSs and the access layer is expressed using the pseudocode in~\refalgo{algo::polyglot}.
   The polyglot access management layer is the entry point for query processing over stored data, providing a unified query interface for accessing the underlying KGs.
   This layer receives the incoming SPARQL queries and labels each query based on its characteristics into three target taxonomies, namely: subject-subject, subject-object, and tree-like (line 1 in \refalgo{algo::polyglot}).
   The labels are used for routing each query to one (or more than one) of the employed DMSs to increase optimal execution.
   We defer details of the labelling approach to~\refsec{sec:label}.
   After determining which DMS should be used for each query (line 2 in \refalgo{algo::polyglot}), a suitable JIT query translation may be needed for further query execution (line 3-5 in \refalgo{algo::polyglot}).
   For example, the incoming query is written in SPARQL, but it is labeled to be run using MongoDB which cannot support SPARQL directly as an input query language.
   In this case, the incoming query needs to be translated JIT into an equivalent JavaScript-like query, MQL, to be executed over MongoDB.
   Handling this JIT translation process is one of the responsibilities of the access layer.
   The following sections describe the detail of each step.

\begin{algorithm}[htbp]
\DontPrintSemicolon
\SetAlgoLined
\KwInput{SPARQL queries (applications' workload)}
 \Initialization{Let q be an incoming SPARQL query}
 
  $qLabel \longleftarrow$ \textbf{Query\_labeling}($q$)\;
 $DMS \longleftarrow$ \textbf{DMS\_select}($qLabel$)\;
   \If{ $DMS == MongoDB$}{
    $q \longleftarrow$ \textbf{Translate\_MQL}($q$)\;
  }
 $qResult \longleftarrow$ \textbf{Route\_execute}($q$)\;
 \textbf{Return\_result}($qResult$)\;
 
 \caption{SymphonyDB KG query processing}
 \label{algo::polyglot}
\end{algorithm}

\subsection{Database Management System Layer}
\label{sec:DMS}

    Selection of adequate DMSs is vital in maximizing the power of a multi-database system. In our prototype, RDF-3X was selected as a candidate as it is an open-source system widely used in a range of studies such as~\cite{watdiv}.
    One vital property is that RDF-3X creates exhaustive indexes on all permutations of triples along with their binary and unary projections.
    Its query processor is designed to aggressively leverage cache-aware hash and merge joins.
    Virtuoso was selected since it is already employed as the DMS of choice for a broad range of KGs, for example the Linked Data for the Life Sciences project\footnote{Available online: \url{https://bio2rdf.org/sparql}}.
    Virtuoso's physical design is based on a relational table with three columns\footnote{In case of loading named graphs, another column is added, called C.} for S, P, and O (S: Subject, P: Predicate, and O: Object) and carries multiple bitmap indexes over that table to provide a number of different access paths.
    Most recently, Virtuoso added columnar projections to minimize the on-disk footprint associated with RDF data storage. Blazegraph\footnote{It is alleged that the Amazon Neptune is based on Blazegraph.} was selected since it is the DMS behind Wikidata, i.e., a KG constructed from the content of Wikimedia sister projects including Wikipedia, Wikivoyage, Wiktionary, and Wikisource.
    Blazegraph's physical design is based on \mbox{B+trees} to store KGs in the form of ordered data.
    Blazegraph typically uses the following three indexes: SPO, POS, and OSP. MongoDB was selected as a representative document-store.
    Its efficacy for executing queries over KGs has not been researched extensively but some academic prototypes such as~\cite{Jignesh,frankPhdThesis,tomaszuk2010document} have already shown document-stores efficacy in similar contexts.

\subsection{Query Labeling and Execution}
\label{sec:label}

\begin{figure}[hbt]
\centering\includegraphics[width=0.45\textwidth]{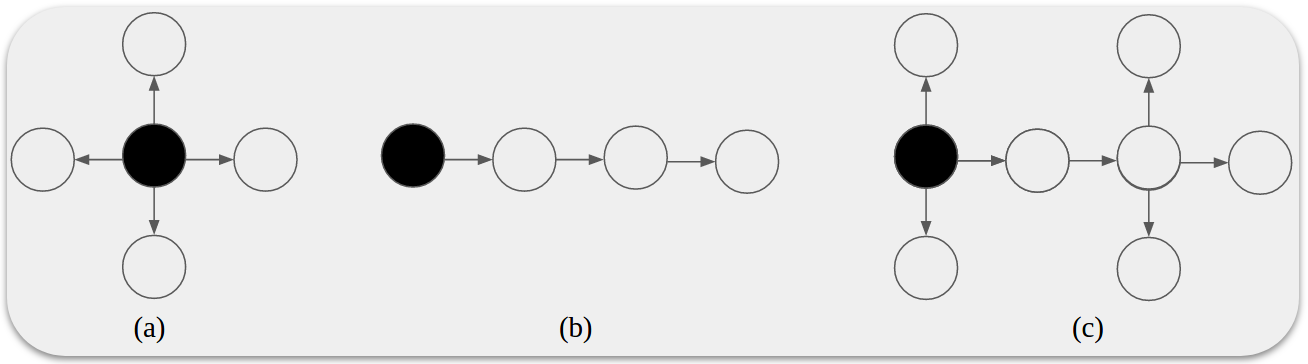}
\caption{(a) Subject-subject query pattern (b) Subject-object query pattern (c) Tree-like query pattern}
\label{fig::qlabel}
\end{figure}

Each SPARQL query can be viewed as a directed graph where nodes are formed by the subjects and objects of the query's triple patterns and edges are the properties of these patterns. Based on this, each SPARQL query can be classified into shape-specific categories.
At this stage, we confine our focus to the following query types: subject-subject, subject-object, and tree-like queries (more details can be found in~\refsec{sec::background}). Examples of these types are shown in~\reffig{fig::qlabel} where each node represents a variable connected through predicates as edges forming a graph-like structure.

In general, each query has at least one source variable as shown in \reffig{fig::qlabel} using nodes represented with solid black colors. Any node with both incoming and outgoing edges represents a shared variable of a query. Query patterns are typically recognizable by the position of shared variables in a query.

 Inspired by~\cite{peter-heuristic}, we utilize a heuristic-based approach to exploit the syntactic and the structural variations of patterns in a given SPARQL query in order to label it.
    On this foundation, SymphonyDB begins by finding the source variable of a given query, looking for all immediate neighbor nodes, or shared variables, with one edge distance away.
    From there, it then iteratively visits nodes further away until all nodes are visited, using a queue data structure to store visited nodes at each stage.
    Upon finishing the traversal, the query will then be labeled according to the characteristics of its variables and graph.
    A subject-subject label is applied if all nodes are only immediate neighbors of the source variable with one edge distance away (\mbox{\reffig{fig::qlabel} (a)}).
    A subject-object label, however, follows a pattern where there is just one outgoing edge from a node at each stage (starting from the source variable) where there is a final node with no outgoing edge as depicted in \mbox{\reffig{fig::qlabel} (b)}.
    Finally, queries that contain a combination of both patterns are labeled as tree-like (\mbox{\reffig{fig::qlabel} (c)}).
    
    In addition to the query patterns, SymphonyDB checks the existence of modifiers as well, where modifiers are keywords such as ``\texttt{LIMIT}'' that are recognizable by the query parser and lexical analyzer implemented in SymphonyDB.

    The following heuristics are then used to select one or more of the integrated DMSs.
    Queries are routed to both MongoDB and RDF-3X if the following characteristics are present: (1)~it contains only a single triple pattern, (2)~it is a query with subject-subject joins, (3)~it contains no modifiers in the query, and (4)~it contains no optional patterns.
    Alternatively, if a query contains subject-object joins, it is routed to Blazegraph, and finally all other queries (tree-like queries with or without optional patterns) are routed to Virtuoso.

    As an intermediate step, any queries being routed to MongoDB must run through the JIT query translation to be translated from SPARQL to MQL. This is explained in the following section.

\subsection{Polyglot Access Management: Query Translation}
\label{sec:translate}

\begin{figure}[h]
\centering\includegraphics[width=0.45\textwidth]{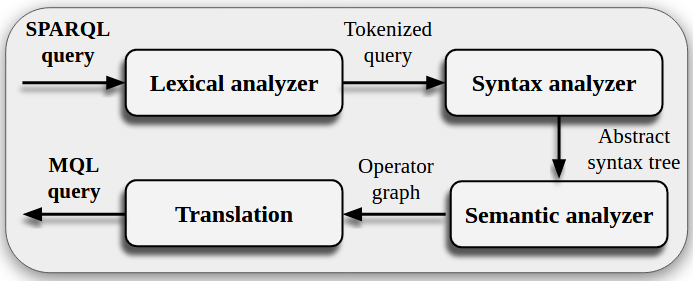}
\caption{The query translation logic flow}
\label{fig::translation}
\end{figure}

Query translation provides the extensibility for SymphonyDB to access various DMSs with differing query languages.
Currently, the only integrated DMS that requires this functionality is MongoDB as it provides its own query language MQL, which is a Javascript-like, object-oriented imperative language.
This contrasts with SPARQL, a domain-specific declarative language~\cite{tomaszuk2010document}, which is not supported natively by MongoDB, however, it is feasible to map SPARQL to MQL in most cases.
\reffig{fig::translation} depicts the logic flow of SymphonyDB's JIT query translation, showing that each query is analyzed lexically and tokenized based on the SPARQL query syntax.
The lexical analyzer dissects the SPARQL query into logical units of one or more characters that have a shared meaning, often referred to as \textit{tokens}.
For instance, ``\texttt{WHERE}'' is a token representing a keywords, whereas ``\texttt{.}'' is an identifier and ``\texttt{=}'' is a sign.
In parallel, it parses each query with regard to the grammatical description of the SPARQL language to generate the corresponding syntax tree.
The semantic analyzer then produces an \textit{operator graph} containing information about projection variables, join patterns, conditions and modifiers.
Finally, once the semantic analysis has completed, heuristic techniques are used to map the operator graph to MQL and translate the query.

    Similar to~\cite{tomaszuk2010document,frankPhdThesis,frankMappping}, SymphonyDB maps each SPARQL to MQL using a collection of rules.
    \reftab{table:rules1} shows SPARQL expressions and their equivalent MQL query string, along with additional set of rules to map SPARQL query patterns to MQL illustrated in~\reftab{table:rules2}.
    For example, to translate subject-subject join queries, SymphonyDB uses the ``\texttt{\$match}'' aggregation pipeline operator of MongoDB to filters documents and pass a subset of the documents that match the specified condition(s) to the next pipeline stage.
    It also uses ``\texttt{\$lookup}'' aggregation pipeline operator of MongoDB to translate joins.

\begin{table}[h]
\caption{SPARQL expressions representation and their equivalent MQL expressions}
\centering
  \begin{tabular}{l | l } 
   \toprule
 SPARQL & MQL  \\ [0.7ex] 
 \hline\hline
 Exists ($<$e1$>$) & $<$e1$>$:\{\$exists:true\}  \\  
 Not Exists ($<$e1$>$) & $<$e1$>$:\{\$exists:false\}  \\ 
 ($<$e1$>$ \&\& $<$e2$>$  ) & \{\$and:[\{$<$e1$>$\},\{$<$e2$>$\}] \\
 ($<$e1$>$ $||$ $<$e2$>$  ) & \{\$or:[\{$<$e1$>$\},\{$<$e2$>$\}] \\
 !($<$e1$>$) & \{\$not:\{$<$e1$>$\}  \\  
 ($<$e1$>$ = $<$e2$>$  ) & \{\$eq:[\{$<$e1$>$\},\{$<$e2$>$\}] \\
 ($<$e1$>$ != $<$e2$>$  ) & \{\$ne:[\{$<$e1$>$\},\{$<$e2$>$\}] \\
 ($<$e1$>$ $>$ $<$e2$>$  ) & \{\$gt:[\{$<$e1$>$\},\{$<$e2$>$\}] \\
 ($<$e1$>$ $>=$ $<$e2$>$  ) & \{\$gte:[\{$<$e1$>$\},\{$<$e2$>$\}] \\
 ($<$e1$>$ $<$ $<$e2$>$  ) & \{\$lt:[\{$<$e1$>$\},\{$<$e2$>$\}] \\
 ($<$e1$>$ $<=$ $<$e2$>$  ) & \{\$lte:[\{$<$e1$>$\},\{$<$e2$>$\}] \\
 \end{tabular}
\label{table:rules1}
\end{table}

\begin{table*}
\caption{Sample rules used to translate different join patterns of SPARQL queries to their equivalent MQL query string}
\resizebox{0.9\textwidth}{!}{
\centering
\begin{tabularx}{\textwidth}{l|l|l}
\cmidrule(lr){1-3}
 \backslashbox{Pattern}{Query}& SPARQL (triple patterns) & MQL (aggregate pipeline)\\ 
 \cmidrule(lr){1-3} \cmidrule(lr){1-3}
\multicolumn{1}{ c|  }{\multirow{3}{*}{Single Triple Pattern} } & 
\multicolumn{1}{ l| }{``subject''  ``predicate''  ``object''} & \{\$match:\{ subject\_id:``subject'', ``predicate'':``object''\}\}\\ 
\multicolumn{1}{ c|  }{}                        &
\multicolumn{1}{ l| }{?subject  ``predicate''  ``object''} & \{\$match:\{ subject\_id:\{\$exists:true\}, ``predicate'':``object''\}\}\\ 
\multicolumn{1}{ c | }{}                        &
\multicolumn{1}{ l| }{``subject''  ``predicate''  ?object} &  \{\$match:\{ subject\_id:``subject'', ``predicate'':\{\$exists:true\}\}\}\\ 

\cmidrule{1-3}  \cmidrule{1-3}
\multicolumn{1}{ c|  }{\multirow{9}{*}{Subject-subject join} } &
\multicolumn{1}{ l| }{\{?subject  ``predicate1''  ``object1'' .} & \{\$match:\{ subject\_id:\{\$exists:true\}, ``predicate1'':``object1'',``predicate2'':``object2''\}\} \\ 
\multicolumn{1}{ c|  }{}                        &
\multicolumn{1}{ l| }{~?subject  ``predicate2''  ``object2'' .\}} & \\ 
\multicolumn{1}{ c | }{}                        &
\multicolumn{1}{ l| }{} &  \\ 
\multicolumn{1}{ c | }{}                        &

\multicolumn{1}{ l| }{\{?subject  ``predicate1''  ?object1~ .} & \{\$match:\{ subject\_id:\{\$exists:true\}, ``predicate1'':\{\$exists:true\},``predicate2'':``object2''\}\} \\ 
\multicolumn{1}{ c|  }{}                        &
\multicolumn{1}{ l| }{~?subject  ``predicate2''  ``object2'' .\}} & \\ 
\multicolumn{1}{ c | }{}                        &
\multicolumn{1}{ l| }{} &  \\ 
\multicolumn{1}{ c | }{}                        &

\multicolumn{1}{ l| }{\{``subject''  ``predicate1''  ``object1'' .} & \{\$match:\{ subject\_id:``subject'', ``predicate1'':\{\$exists:true\},``predicate2'':``object2''\}\} \\ 
\multicolumn{1}{ c|  }{}                        &
\multicolumn{1}{ l| }{~``subject''  ``predicate2''  ``object2'' .\}} & \\ 
\multicolumn{1}{ c | }{}                        &
\multicolumn{1}{ l| }{} &  \\ 


\cmidrule{1-3}  \cmidrule{1-3}


\multicolumn{1}{ c|  }{\multirow{12}{*}{Subject-object join} } &
\multicolumn{1}{ l| }{\{?subject  ``predicate1''  ?object1~ .} & \{\{\$match:\{ subject\_id:\{\$exists:true\}\}\},\\ 
\multicolumn{1}{ c|  }{}                        &
\multicolumn{1}{ l| }{~?object1  ``predicate2''  ``object2'' .\}} & \{\$lookup:\{ from: ``colc\_name'', localField: ``predicate1'', foreignField: ``subject\_id'', as: ``join\_field''\}\},\\ 
\multicolumn{1}{ c | }{}                        &
\multicolumn{1}{ l| }{} & \{\$match:\{ ``join\_field.predicate2'':``object2''\}\}\}\\ 
\multicolumn{1}{ c | }{}                        &
\multicolumn{1}{ l| }{} &  \\ 
\multicolumn{1}{ c | }{}                        &

\multicolumn{1}{ l| }{\{?subject  ``predicate1''  ?object1~ .} & \{\{\$match:\{ subject\_id:\{\$exists:true\}\}\},\\ 
\multicolumn{1}{ c|  }{}                        &
\multicolumn{1}{ l| }{~?object1  ``predicate2''  ?object2 .\}} & \{\$lookup:\{ from: ``colc\_name'', localField: ``predicate1'', foreignField: ``subject\_id'', as: ``join\_field''\}\},\\ 
\multicolumn{1}{ c | }{}                        &
\multicolumn{1}{ l| }{} & \{\$match:\{ ``join\_field.predicate2'':\{\$exists:true\}\}\}\}\\ 
\multicolumn{1}{ c | }{}                        &
\multicolumn{1}{ l| }{} &  \\ 
\multicolumn{1}{ c | }{}                        &

\multicolumn{1}{ l| }{\{``subject''  ``predicate1''  ?object1~ .} & \{\{\$match:\{ subject\_id:``subject''\}\},\\ 
\multicolumn{1}{ c|  }{}                        &
\multicolumn{1}{ l| }{~?object1  ``predicate2''  ?object2 .\}} & \{\$lookup:\{ from: ``colc\_name'', localField: ``predicate1'', foreignField: ``subject\_id'', as: ``join\_field''\}\},\\ 
\multicolumn{1}{ c | }{}                        &
\multicolumn{1}{ l| }{} & \{\$match:\{ ``join\_field.predicate2'':\{\$exists:true\}\}\}\}\\ 
\multicolumn{1}{ c | }{}                        &
\multicolumn{1}{ l| }{} &  \\ 


\cmidrule{1-3}  \cmidrule{1-3}


\multicolumn{1}{ c|  }{\multirow{12}{*}{Tree-like join} } &
\multicolumn{1}{ l| }{\{?subject  ``predicate1''  ?object1~ .} & \{\{\$match:\{ subject\_id:\{\$exists:true\}, ``predicate1'':\{\$exists:true\}, ``predicate2'':\{\$exists:true\}  \}\},\\ 
\multicolumn{1}{ c|  }{}                        &
\multicolumn{1}{ l| }{~?subject  ``predicate2''  ?object2~ .} &  \{\$lookup:\{ from: ``colc\_name'', localField: ``predicate2'', foreignField: ``subject\_id'', as: ``join\_field''\}\},\\ 
\multicolumn{1}{ c | }{}                        &
\multicolumn{1}{ l| }{~?object2  ``predicate3''  ``object3'' .\}} & \{\$match:\{ ``join\_field.predicate3'':``object3''\}\}\}\\ 
\multicolumn{1}{ c | }{}                        &
\multicolumn{1}{ l| }{} & \\ 
\multicolumn{1}{ c | }{}                        &

\multicolumn{1}{ l| }{\{``subject''  ``predicate1''  ``object1''~ .} & \{\{\$match:\{ subject\_id:``subject'', ``predicate1'':``object1'', ``predicate2'':\{\$exists:true\}  \}\},\\ 
\multicolumn{1}{ c|  }{}                        &
\multicolumn{1}{ l| }{~``subject''  ``predicate2''  ?object2~ .} &  \{\$lookup:\{ from: ``colc\_name'', localField: ``predicate2'', foreignField: ``subject\_id'', as: ``join\_field''\}\},\\ 
\multicolumn{1}{ c | }{}                        &
\multicolumn{1}{ l| }{~?object2  ``predicate3''  ?object3 .\}} & \{\$match:\{ ``join\_field.predicate3'':\{\$exists:true\}\}\}\}\\ 
\multicolumn{1}{ c | }{}                        &
\multicolumn{1}{ l| }{} & \\ 
\multicolumn{1}{ c | }{}                        &

\multicolumn{1}{ l| }{\{``subject''  ``predicate1''  ?object1~ .} & \{\{\$match:\{ subject\_id:``subject'', ``predicate1'':\{\$exists:true\}  \}\},\\ 
\multicolumn{1}{ c|  }{}                        &
\multicolumn{1}{ l| }{~?object1  ``predicate2''  ?object2~ .} &  \{\$lookup:\{ from: ``colc\_name'', localField: ``predicate1'', foreignField: ``subject\_id'', as: ``join\_field''\}\},\\ 
\multicolumn{1}{ c | }{}                        &
\multicolumn{1}{ l| }{~?object1  ``predicate3''  ``object3'' .\}} & \{\$match:\{ ``join\_field.predicate2'':\{\$exists:true\}, ``join\_field.predicate3'':``object3''\}\}\}\\ 
\multicolumn{1}{ c | }{}                        &
\multicolumn{1}{ l| }{} & \\ 

\cmidrule{1-3}  \cmidrule{1-3}

\end{tabularx}
}
\label{table:rules2}
\end{table*}

\section{Experimental Evaluation}
\label{sec::Experimental-Setting}

    In this section, we report the experimental setup and details of the KG benchmark datasets that are used in the experimental evaluation.
    This includes detailed information about DMSs' configuration, indexing, data loading process as well as our computational platform.
    The query performance of SymphonyDB and a range of DMSs are evaluated and presented below.

\subsection{Evaluation Datasets and Queries}

    We select four well-known KG datasets with a collection of relevant queries that are publicly available, where a number have been used in previous studies~\cite{cellcycle,biobench}.
    The datasets are as follows.
    \textbf{Allie}\footnote{\url{http://allie.dbcls.jp/}} is a KG surrounding life sciences, containing abbreviations and long forms utilized within the field.
    \textbf{Cellcycle}\footnote{\url{ftp://ftp.dbcls.jp/togordf/bmtoyama/cellcycle/}} contains orthology relations for proteins consiting of ten sub-graphs constituting the cell cycle.
    In our experiments, however, we integrated all ten sub-graphs into a single KG dataset without modifying any content.
    \textbf{DrugBank}\footnote{\url{https://download.bio2rdf.org/files/current/drugbank/drugbank.html}} contains bioinformatics and chemoinformatics resources which include detailed drug (chemical, pharmacological, pharmaceutical, etc.) and comprehensive drug targets (sequence, structure and pathway information) in the dataset.
    \textbf{LinkedSPL\footnote{\url{https://download.bio2rdf.org/files/current/linkedspl/linkedspl.html}}} includes all sections of FDA-approved prescriptions and over-the-counter drug package inserts from DailyMed. \reftab{table::c5:kgs} depicts the statistical information related to the above KGs.

\begin{table}[tbp]
\centering
\caption{KGs that were used to run the experiments}
\begin{tabularx}{\linewidth}{l r r r r}
\toprule
\multirow{2}{*}{KG}& \multicolumn{4}{c}{Statistics} \\
\cmidrule{2-5}
& Sub. (\#) & Pre. (\#)& Obj. (\#) & Triples (\#)\\
 \midrule
 \midrule
 Allie & 19,227,252 & 26 & 20,280,252 & 94,404,806 \\ 
 Cellcycle & 21,745 & 18 & 142,812 & 322,751\\
 DrugBank & 19,693 & 119 & 276,142 & 517,023 \\
LinkedSPL & 59,776 & 104 & 719,446 & 2,174,579 \\
 \bottomrule
\end{tabularx}

\label{table::c5:kgs}
\end{table}

\begin{table*}
\caption{Types of the queries. $SS^{a*}$: Subject-subject join, $SO^{b*}$: Subject-object join, $Co^{c*}$: combination of $SS$ and $SO$, $OPT^{d*}$: Optional pattern, $Fil^{e*}$: Filter, $ORD^{f*}$: Order by, $Lim^{g*}$: Limit, $OFF^{h*}$: Offset, $STP^{i*}$: Single triple pattern (no join)}

\centering
\begin{tabular}{cccccccccccc}
\toprule
 \multirow{3}{*}{Benchmark}& \multicolumn{11}{c}{Types} \\
 \cmidrule{3-12}
  & Query & $SS^{a*}$ & $SO^{b*}$ & $Co^{c*}$ & $OPT^{d*}$ & Selective & $Fil^{e*}$ & $ORD^{f*}$ & $Lim^{g*}$ & $OFF^{h*}$ & $STP^{i*}$\\
 \hline\hline
\multicolumn{1}{ c  }{\multirow{5}{*}{Allie} } &
\multicolumn{1}{ c }{Q1} &  &  &  & & & & & & &  \cmark\\ 
\multicolumn{1}{ c  }{}                        &
\multicolumn{1}{ c }{Q2} &  &  &  & & & \cmark & & & &  \cmark\\ 
\multicolumn{1}{ c  }{}                        &
\multicolumn{1}{ c }{Q3} & \cmark &  &  & & &  & & & & \\ 
\multicolumn{1}{ c  }{}                        &
\multicolumn{1}{ c }{Q4} &  & \cmark  &  & & &  & & \cmark & & \\ 
\multicolumn{1}{ c  }{}                        &
\multicolumn{1}{ c }{Q5} & \cmark  & &  & & &  & \cmark & \cmark & & \\ \cline{1-12}
\hline

\multicolumn{1}{ c }{\multirow{6}{*}{Cellcycle} } &
\multicolumn{1}{ c }{Q1} &  &  & \cmark & & & & & & &  \\ 
\multicolumn{1}{ c }{}                        &
\multicolumn{1}{ c }{Q2} &  &  & \cmark & \cmark & & & & & &  \\ 
\multicolumn{1}{ c }{}                        &
\multicolumn{1}{ c }{Q3} &  &  & \cmark &  & \cmark & & & & &  \\ 
\multicolumn{1}{ c }{}                        &
\multicolumn{1}{ c }{Q4} &  &  & \cmark &  & & & & & &  \\ 
\multicolumn{1}{ c }{}                        &
\multicolumn{1}{ c }{Q5} & \cmark  &  &  & \cmark & & & & & &  \\ \cline{1-12}

\hline
\multicolumn{1}{ c  }{\multirow{5}{*}{DrugBank} } &
\multicolumn{1}{ c }{Q1} & \cmark &  &  &  \cmark & & & & \cmark & &  \\ 
\multicolumn{1}{ c  }{}                        &
\multicolumn{1}{ c }{Q2} & \cmark &  &  & \cmark & &  & \cmark &\cmark &\cmark & \\ 
\multicolumn{1}{ c  }{}                        &
\multicolumn{1}{ c }{Q3} &  &  & \cmark & & &  & & & & \\ 
\multicolumn{1}{ c  }{}                        &
\multicolumn{1}{ c }{Q4} &  &   & \cmark & & &  & & & & \\ 
\multicolumn{1}{ c  }{}                        &
\multicolumn{1}{ c }{Q5} &   &  & \cmark & & &  &  & \cmark & & \\ \cline{1-12}

\hline
\multicolumn{1}{ c  }{\multirow{2}{*}{LinkedSPL} } &
\multicolumn{1}{ c }{Q1} &  \cmark &   &  & & &  & &\cmark & & \\ 
\multicolumn{1}{ c  }{}                        &
\multicolumn{1}{ c }{Q2} &   &  & \cmark & & &  &\cmark  & \cmark &\cmark & \\ \cline{1-12}

\end{tabular}

\label{table::queries}
\end{table*}

We selected 17 representative queries\footnote{Available through \url{https://github.com/m-salehpour/SymphonyDB}}. \reftab{table::queries} shows the classification of the 17 queries. A range of these queries have also been used in previous studies such as~\cite{biobench,cellcycle,saleem}.

\subsection{Evaluation Platform} 

\noindent \textbf{Computational Environment.} Our benchmark system was a physical machine with a 3.4GHz Core i7-3770 Intel processor, running Ubuntu Linux (kernel version: 4.15.0-91-generic), with 16GB of main memory, 8 cores, 256K L2 cache, 1TB instance storage capacity.

\noindent \textbf{Data Management Systems (DMSs).} Our DMSs: (1) Virtuoso (version 07.20.3230), (2) Blazegraph (version 2.1.6), RDF-3X (version 0.3.8), and MongoDB (version 4.2.3). All or some of these DMSs have also been used in previous studies such as~\cite{watdiv,ISWC2013,Medha2,bsbm,saleem,biobench}.
We configured these DMSs based on their vendors' official recommendations.
We did not change the default indexing scheme of the DMSs since they usually create exhaustive indexes over all permutations of RDF triples. Note that creating alternate indexing schemes is feasible but will not be generally needed.\footnote{\url{http://docs.openlinksw.com/virtuoso/rdfperfrdfscheme}} For MongoDB, we created indexes on those name/value pairs of the JSON representations that were representatives of subjects and predicates. We loaded the RDF/N-Triples format of KGs into DMSs by using their native bulk loader functions. We converted the KG datasets from RDF/N-Triples syntax to JSON-LD using a parser designed and developed as part of this research\footnote{Available through \url{https://github.com/m-salehpour/SymphonyDB}} to load them into MongoDB.

\noindent \textbf{Measurement.} 
The query times for cold-cache are reported below.
We dropped the cache using the following commands: \texttt{echo 3 > /proc/sys/vm/drop\_caches} and \mbox{\texttt{swapoff -a}}. The output of each query was verified to ensure that output results were correct and consistent across the different DMSs.

\begin{figure*}
        \centering
        \subfigure[]{\label{result:c5:a}\includegraphics[width=.375\textwidth]{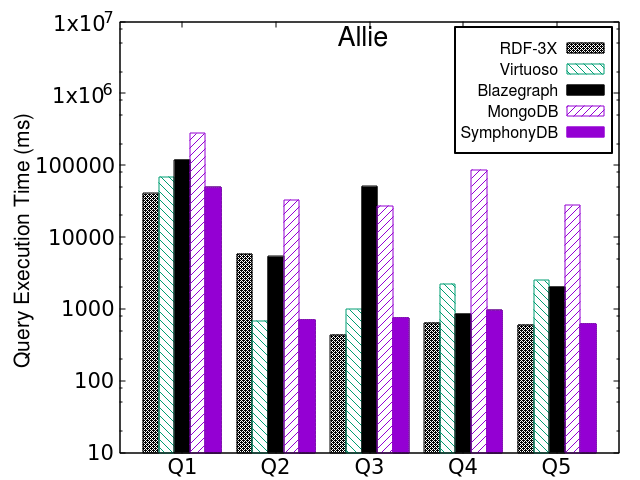}}
        \subfigure[]{\label{result:c5:b}\includegraphics[width=.375\textwidth]{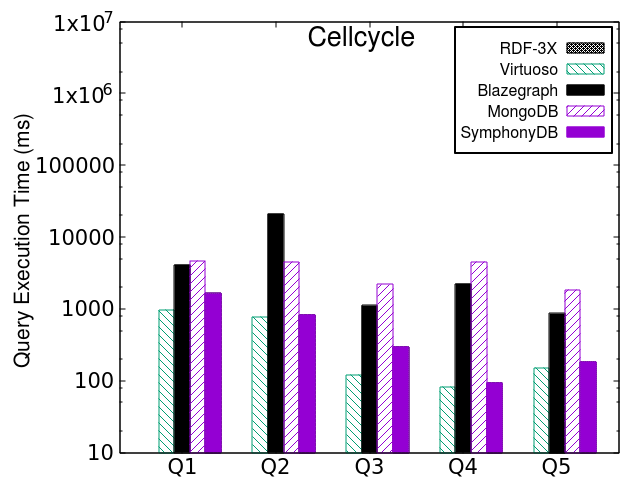}}
        \subfigure[]{\label{result:c5:c}\includegraphics[width=.375\textwidth]{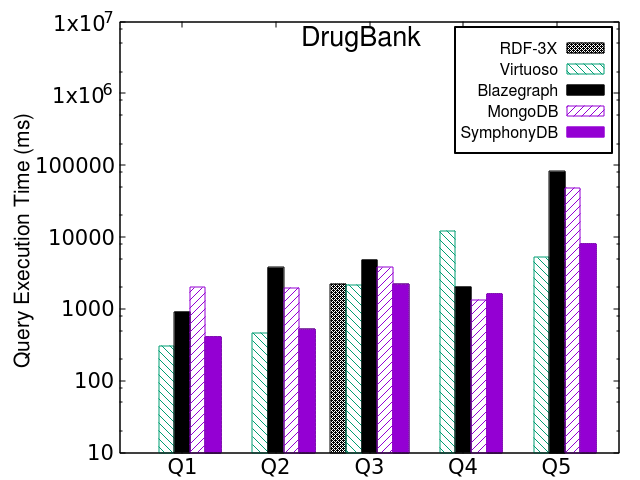}}
        \subfigure[]{\label{result:c5:d}\includegraphics[width=.375\textwidth]{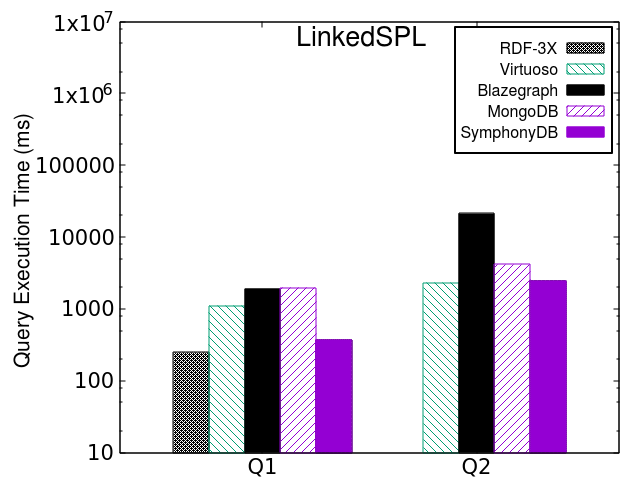}}
    \caption{ The execution times of different queries against each KG. The $X$ axis shows different queries. The $Y$ axis shows the execution time of each query in milliseconds (\textbf{log scale}). No value is shown when a query is not supported by a DMS (e.g., RDF-3X does not support queries of Cellcycle KG) or the returned result is different with others (e.g., Blazegraph's result for Allie-Q5).\label{fig:c5:result}} 
        
\end{figure*}


\section{Results}
\label{sec::c5:results}
The query execution times over the KGs are presented in \reffig{fig:c5:result} where the $X$ axis shows the different queries and the $Y$ axis shows the execution times in milliseconds (\textbf{log scale}). These results suggest that RDF-3X offers several orders of magnitude performance advantages over others for queries with a single triple pattern (i.e., no join) and less complex triple patterns (e.g., no optional or complex filtering patterns) such as Allie-Q1 and Allie-Q3-Allie-Q5. However, this DMS could not execute Allie-Q2 as fast as others since this query contains a filtering pattern. Note that RDF-3X could not execute queries with complex triple patterns or offset modifiers (e.g.,~\reffig{result:c5:b}). In these cases, no value is shown.
Virtuoso exhibits around one order of magnitude better performance to run complex queries containing a combination of subject-subject and subject-object joins.
As compared to Virtuoso, Blazegraph showed relatively better performance to execute subject-object join queries like Allie-Q4.
MongoDB as a document-store could execute all the queries. For subject-subject join queries like DrugBank-Q1, DrugBank-Q2, and LinkedSPL-Q1, its performance is comparable with others.

Our results indicate that SymphonyDB performs \textit{consistently} across different datasets. Its performance is almost equal to the fastest DMSs in all cases. More specifically, SymphonyDB is consistently the second-best DMS (with a negligible difference as compared to the best DMS) for executing different queries.

\subsection{Discussion}

In this section, we provide a discussion on the experimental results, detailing further insight into the results, and providing key takeaways for SymphonyDB.

\subsubsection{Analysis}
\label{sec::analysis}

Although the results indicate SymphonyDB's consistency across ranging query types, various factors contribute to the performance differences of SymphonyDB with other DMSs.
SymphonyDB labels queries based on strict characteristics and heuristics, such as the number of triple patterns, modifiers, optional patterns, and a number of query join patterns.
This classification forms the basis for which underlying DMS is selected to route the given query to.
For instance, Allie KG queries were routed to RDF-3X and MongoDB (after translation) as two of these queries contained the \textit{single triple pattern}, shown in Allie-Q1 and Allie-Q2, and others contained subject-subject patterns with no modifiers or optional patterns.
This analysis is essential in routing queries to the most optimal DMS, even though it imposes minor overheads.

The overhead of query labeling increases with the need for query translation if routing to a DMS that requires this functionality.
It is plausible that this overhead influenced the performance of SymphonyDB, justifying that the performance difference between SymphonyDB and the best execution time for each query.
However, the significance of these overheads can be overlooked due to the consistency in performance across a range of query types exhibited by SymphonyDB.
Thus, the overheads observed as a result of labeling and translation are viewed as a small trade-off for added consistency in query execution performance.

\subsubsection{Limitations}
\label{sec::Discussion}

    Although performance improvements were observed during experimental conditions, there are a number of hindrances experienced by multi-database environments that still pose challenges to increased performance.
    Limitations experienced by SymphonyDB include:

        \begin{enumerate}[itemsep=0pt,label=\roman*)]
            \item \textbf{Replication of KG datasets} --- As multi-database systems employ multiple DMSs, it requires the datasets replicated on each system. The number of replications is determined by the number of DMSs utilized in the underlying layers, e.g. this number is equal to four for SymphonyDB.
            For write-heavy applications, this replication can lead to increased latency during write operations and therefore decreased write performance.
            However, most KG applications tend to be \textit{read-mostly} if not \textit{read-only}~\cite{Hexastore,RDF3x}. Thus, write latency is not a concern in most use cases.

            \item \textbf{Efficiency of translations} --- All SPARQL queries may not be translated to (efficient) MQL queries due to the dissimilarity between the expressiveness of SPARQL and MQL~\cite{frankPhdThesis}. For instance, triple patterns whose predicates are replaced by variables could not be translated into an efficient query for being executed over MongoDB (in most cases). In this research, we did not have such queries and we carefully checked to ensure that our JIT query translation can produce correct and efficient MQL queries for the benchmark SPARQL queries.
            However, future work entails further optimization and improvements on the translation to ensure optimality of the query.
        \end{enumerate}

\section{Related Work}
\label{sec::related_work}

    Over the last few years, there has been growing interest in multi-database solutions resulting in the research and development of open-source platforms, such as Apache Beam and Drill, as well as academic prototypes~\cite{bigdawg}.
    In general, these proposals utilize a model consisting of multiple DMSs but require input from expert users to decide which specific DMS meets the requirement for a given application or query set.
    For example, \cite{bigdawg} presents two commands, namely \texttt{scope} and \texttt{cast} which provides a user with information to select the most appropriate DMS for the query being analysed.
    Recent works~\cite{rheemshort,Musketeershort,f1} present parallel cross-platform data processing systems to decouple application interaction from underlying platforms.
    These systems follow a process that splits each given query into sub-queries, executing them on multiple platforms simultaneously to minimize the overall runtime.
    Although providing a speedup, it is unclear, however, how much of the performance gain comes from minimizing inter-platform communication overheads by taking advantage of data locality for sub-query processing.
    Various proposals take alternative approaches, presenting cross-platform \textit{stream processing}~\cite{lim2013} or building dynamic workload management through adaptable architecture design.
    These various proposals show that current multi-database solutions primarily focus on applications such as data integration, ETL, machine learning, stream processing, etc. and pay little attention to employ multiple DMSs for high-performance KG query processing.

\section{Conclusion}
\label{sec::conclusion}
The increases in the heterogeneity of KG datasets have triggered the development of a range of DMSs broadly classified as key-value, document, columnar, and graph stores in addition to the relational. There exists no single DMS that meets the diverse performance requirements of KG query processing efficiently. In this paper, we have addressed some of the critical performance challenges associated with current DMSs in the context of KGs by proposing an architecture that can achieve polygloty for KG query processing supported by a unified access management layer. This approach potentially can be extendable to the non-monolithic conception of database processing in which the different components such as file systems, index structures compression, query processing engines, concurrency, consistency modules, etc. are made available in the cloud and communicate through high-performance networks~\cite{kossmannshort,Aurorashort}. Further steps will also include efforts to minimize the amount of data replications without negatively affecting the robustness and performance.

\balance

\small
\bibliographystyle{abbrv}
\bibliography{main}

\end{document}